\documentclass[twocolumn,showpacs,superscriptaddress,pra]{revtex4}

\pdfoutput=1
\usepackage{amsmath, amsthm, amssymb}
\usepackage{graphicx}
\usepackage{dcolumn}
\usepackage{bm}
\usepackage{color} 

\newcommand{\ket}[1]{| #1 \rangle}
\newcommand{\bra}[1]{\langle #1 |}

\newcommand{\tr}{\mathrm{Tr}}

\newcommand{\eg}{{\it{e.g.~}}}
\newcommand{\ie}{{\it{i.e.~}}}
\newcommand{\etal}{{\it{et al.}}}

\begin{document}

\title{Macroscopic bound entanglement in thermal graph states}

\author{D. Cavalcanti}
\affiliation{ICFO-Institut de Ciencies Fotoniques, Mediterranean
Technology Park, 08860 Castelldefels (Barcelona), Spain}

\author{L. Aolita}
\affiliation{ICFO-Institut de Ciencies Fotoniques, Mediterranean
Technology Park, 08860 Castelldefels (Barcelona), Spain}

\author{A. Ferraro}
\affiliation{ICFO-Institut de Ciencies Fotoniques, Mediterranean
Technology Park, 08860 Castelldefels (Barcelona), Spain}

\author{A. Garc\'ia-Saez}
\affiliation{ICFO-Institut de Ciencies Fotoniques, Mediterranean
Technology Park, 08860 Castelldefels (Barcelona), Spain}

\author{A. Ac\'in}
\affiliation{ICFO-Institut de Ciencies Fotoniques, Mediterranean
Technology Park, 08860 Castelldefels (Barcelona), Spain}
\affiliation{ICREA-Instituci\'o Catalana de Recerca i Estudis
Avan\c cats, Lluis Companys 23, 08010 Barcelona, Spain}

\begin{abstract}
We address the presence of bound entanglement in strongly-interacting spin systems at thermal equilibrium. In particular, we consider thermal graph states composed of an arbitrary number of particles. We show that for a certain range of temperatures no entanglement can be extracted by means of local operations and classical communication, even though the system is still entangled. This is found by harnessing the independence of the entanglement in some bipartitions of such states with the system's size. Specific examples for one- and two-dimensional systems are given.
Our results thus prove the existence of thermal bound entanglement in an arbitrary large spin system with finite-range local interactions.
\end{abstract}


\maketitle

\section{Introduction}

The application of tools recently developed in the context of quantum information theory to
problems  historically native from the area of many-body physics has helped us gain new insight  about collective quantum phenomena \cite{AmicoReview}. In particular,  the characterization of
the entanglement properties of ground and thermal states of strongly-interacting spin Hamiltonians provides a description of these systems from a novel and alternative perspective. Also, in general, it gives us information about how hard it is to simulate their dynamical  and statical properties with classical resources \cite{Vidal}. In addition, such characterization is crucial to determine when these systems can in turn assist as resources in some given quantum-information processing task. For all these reasons it is important to classify and characterize standard many-body
models in terms of their entanglement properties.

\par However, the entanglement characterization in multiparticle systems turns out to be formidably hard. On one hand, the calculation of truly multipartite entanglement measures is in general extremely difficult, even for the pure-state case \cite{PlenioVirmani,HorReview}. On the other hand, for realistic many (and specially macroscopically many)-body systems, the interaction with its surrounding environment can additionally never be neglected and mixed states have to be necessarily taken into account. Unfortunately though, the
characterization of mixed-state entanglement is up to date very poorly developed even for the general bipartite case. One of the most frequent and important type of such interactions -- the one on which we will focus here -- takes place when the system is embedded in a thermal bath at temperature $T$ and reaches thermal equilibrium with it, a process called thermalization. This process typically causes the system to lose quantum coherence, gain entropy and in most times also lose entanglement.

\par An important step forwards in the characterization of mixed-state entanglement was to recognize two different
types of entanglement: distillable and bound entanglement
\cite{Horod}. An entangled state $\rho$ is said to be \emph{distillable}
if it is possible, by means of local operations and classical
communication (LOCC), to obtain from $\rho$ (or, more precisely, several copies thereof) pure-state entanglement. Entangled states
for which this task is impossible are said to be \emph{bound-entangled}. Bound-entangled states were for some time believed to
be useless for quantum information processing. Nevertheless, they are nowadays known to be
useful in some practical situations \cite{Useful
Bound}.

Historically, most examples of bound entangled states have been
provided without relying on any operational recipe (see \eg
Ref.~\cite{HorReview}). Yet this kind of entanglement has been
recently found to arise in natural processes, in particular
thermalization \cite{Oscillators,Toth,Patane, KasKay,Aolita} and
other dynamical decoherence processes \cite{Aolita}. For thermal
states, the presence of bound entanglement was found in chains of
harmonic oscillators in the thermodynamic limit, that is for an
arbitrary number of oscillators \cite{Oscillators}. On the other
hand, in the case of fermionic systems the existence of thermal
bound entanglement has been shown only for small systems of up to
12 spins \cite{Oscillators,Toth,Patane} or for models involving
non-local interactions between an arbitrary number of
spins\cite{KasKay}. The main contribution of the present paper is
thus to show the existence of bound entangled thermal states in
strongly-correlated systems of arbitrary number $N$ of
spin-$\frac{1}{2}$ particles with local interactions, in
particular in the limit $N\longrightarrow\infty$.

\par We show the presence of thermal bound entanglement for the exemplary family of Hamiltonians through which graph states are defined \cite{graph_review}. These Hamiltonians are always frustration-free and typically, depending on the graph, also local (meaning that not all particles interact simultaneously) and of finite-range (meaning that no particle interacts with another one infinitely far away) \cite{Exception}. Graph states constitute an extremely important
family of states from practical and fundamental points of view. They include cluster states, which are resources for universal measurement-based quantum
computation \cite{RausBrie}, codeword states for quantum error correction
\cite{SchWer}, and the well-known GHZ states, that are resources for secure quantum communication
\cite{DurCasBrie-ChenLo}. Moreover, this family of states can be used in quantum non-locality tests
\cite{GHZ,Scarani, GuhneGraph,GuhneCabello}.

\par
The motivation of this work is thus twofold: From a practical point of view, to establish the range of temperatures for which a thermal state of an interesting many-body model proves useful as a resource for some potential quantum information processing task. And, from a fundamental viewpoint, to take this particular case as a concrete example within a broader investigation of the properties
of quantum correlations of strongly-correlated systems undergoing open-system dynamics.

\section{Thermal graph states and dephased graph states}

In this section we define the states under scrutiny and settle the notation. Let us then start by defining a mathematical graph $G\equiv\{\mathcal{V},\mathcal{E}\}$ as the union of the set $\mathcal{V}$ of vertices $i\in\mathcal{V}$ with the set $\mathcal{E}$ of  edges  $\{i,j\}\in
\mathcal{E}$ connecting each vertex $i$ to some other $j$, being $1\leq i,\ j\leq N$. Next, for each graph $G$ we define Hamiltonian $H$ acting on $N$ spin-$1/2$ particles as
\begin{equation}
\label{H}
H=-\frac{1}{2}\sum_{i=1}^N B_i X_i\otimes\bigotimes_{j\in\mathcal{N}_i} Z_{j},
\end{equation}
where $B_i>0$ are arbitrary (strictly positive) coupling strengths in arbitrary units, $X_k$ and $Z_k$ are the usual Pauli operators acting on particle $k$, and $\mathcal{N}_k$ denotes all neighbouring particles of $k$ -- \ie each particle  whose graph representation in $G$ is a vertex $j$ directly connected to $k$ by some edge $\{j,k\}\in
\mathcal{E}$.
Hamiltonian \eqref{H} involves $m$-body interactions, where $m$ is given by the maximum connectivity of the graph $G$. Also, since each and all of the $N$ terms in summation \eqref{H} commute with each other, the eigenstates -- here denoted by $\ket{G_{\mu_1\ ...\ \mu_N}}$, with $\mu_i=0$ or 1, for $1\leq i\leq N$ -- of each local term  are also eigenstates of  the whole summation. This implies that Hamiltonian \eqref{H} is frustration-free, meaning that the ground state --  $\ket{G_{0\ ...\ 0}}$ -- minimizes the energy of each term in $H$.

\par Ground state $\ket{G_{0\ ...\ 0}}$ is the unique (non-degenerate) ground state of Hamiltonian \eqref{H} with eigenenergy $-\frac{1}{2}\sum_{i=1}^N B_i$, whereas states $\ket{G_{\mu_1\ ...\ \mu_N}}$ -- related to the former by the local-unitary transformation $\ket{G_{\mu_1\ ...\ \mu_N}}\equiv\bigotimes_{i=1}^N {Z_{i}}^{\mu_i}\ket{G_{0\ ...\ 0}}$ -- are eigenstates of  \eqref{H} with $\sum_{i=1}^N\mu_i$ excitations and associated eigenenergies $-\frac{1}{2}\sum_{i=1}^N B_i (-1)^{\mu_i}$  \cite{graph_review}. The eigenstates $\ket{G_{\mu_1\ ...\ \mu_N}}$ form a complete orthogonal basis of the $N$-qubit Hilbert space. Since they are local-unitarily related, they all possess exactly the same entanglement properties, extensively studied in Ref.~\cite{graph_review,HeinEisBrie} and references therein. For historical reasons, the ground state $\ket{G_{0\ ...\ 0}}$ has been taken however as the defining state for the so-called graph state. Let us recall that there exists also an alternative operational definition for such graph state: It can be physically produced initializing $N$ qubits in the superposition $\ket{+}=(\ket{0}+\ket{1})/\sqrt{2}$ and subsequently applying control-$Z$ gates $CZ_{ij}=e^{i\frac{\pi}{4}(Z_{i}\otimes Z_{j}-Z_{i}-Z_{j}+\openone)}$ onto each pair of neighboring qubits defined by the graph $G$. Mathematically:
\begin{equation}
\label{graph state}
\ket{G_{0\ ...\ 0}}=\bigotimes_{i=1}^N \bigotimes_{j\in\mathcal{N}_i}CZ_{ij}\bigotimes_{k=1}^N\ket{+}_k.
\end{equation}

We are now in condition to introduce the thermal graph state associated to $G$ as the thermal  state of Hamiltonian \eqref{H}:
\begin{equation}
\label{rho thermal}
\rho_{T}=\frac{e^{-H/ T}}{\tr\big[e^{-H/ T}\big]},
\end{equation}
where  $T$ is the temperature of some bath with which our system of interest has reached thermal equilibrium (Boltzmann's constant is set as unit $k_B\equiv1$ throughout).

\par Alternatively, $\rho_T$ can be defined as a decohered graph state. In order to show that we need to introduce the completely-positive map $\Lambda$, acting on any $N$-qubit density matrix $\rho$ as
\begin{equation}
\label{decoherence}
\Lambda(\rho)\equiv\mathcal{D}_1\otimes ...\ \mathcal{D}_N(\rho),
\end{equation}
as the composition of local, independent channels $\mathcal{D}_i$,
\begin{equation}
\label{dephasing}
\mathcal{D}_i(\rho)=\big(1-\frac{p_i}{2}\big)\rho+\frac{p_i}{2}Z_i \rho Z_i,
\end{equation}
with $0\leq p_i\leq 1$. Local channel $\mathcal{D}_i$ describes the physical process in which, with probability $p_i$, an undesired $\pi$-phase shift is experienced by qubit  $i$ and, with probability $1-p_i$, the system is left  untouched. Such process is present in situations where, with probability $p_i$, there is complete loss of quantum coherence
but without any population exchange. In the context of decoherence, map $\Lambda$ in turn is often referred to as \emph{local} (or \emph{individual})  \emph{dephasing} (or \emph{phase damping}).

\par It was shown in Ref. \cite{Kay} that thermal state \eqref{rho thermal} -- for the particular case of constant couplings $B_i\equiv B$ -- can also be alternatively obtained by individually dephasing graph state \eqref{graph state} with equal probabilities  $p_i\equiv p=\frac{2}{1+e^{B/T}}$. As  shown in Appendix  \ref{App}, this property also holds for general thermal states of Hamiltonian \eqref{H} with arbitrary couplings. Indeed,
\begin{equation}
\label{decoherence1}
\rho_{T}\equiv\Lambda\big(\ket{G_{0\ ...\ 0}}\bra{G_{0\ ...\ 0}}\big),
\end{equation}
with  $\Lambda$ defined according to Eqs. \eqref{decoherence} and \eqref{dephasing}, but with the additional constraint that the local dephasing probabilities satisfy
\begin{equation}
\label{probability}
\frac{p_i}{2}\equiv\frac{1}{1+e^{B_i/T}}.
\end{equation}
The equivalence mathematically expressed in Eqs. \eqref{decoherence1} and \eqref{probability} establishes a very interesting connection between a collective decoherence process [thermalization of systems governed by graph-state Hamiltonians as \eqref{H}] and a local one [individual dephasing of systems initialized in graph states as \eqref{graph state}].

\section{Appearance of multipartite bound entanglement in thermal graph states}

As mentioned before, the characterization of entanglement in multiqubit systems is formidably hard even for pure states. However, a useful tool for the evaluation of the amount of entanglement contained in decohered graph states was developed in Ref. \cite{Cavalcanti} for an important family of decoherence processes (see also Refs. \cite{Kay,HeinDurBrie}). Given a certain multipartition of a graph $G$, the machinery developed in \cite{Cavalcanti} allows to map the calculation of the entanglement of a decohered (mixed) graph state to the average entanglement of several effective systems, constituted only by the so-called {\it boundary qubits} -- the ones lying on the border of the multipartition and having neighbours on the other side of the border. Solving the former problem involves an optimization over a parameter space exponentially large with $N$, whereas solving the latter involves only an optimization over the boundary qubits, a task that requires always exponentially less memory space and usually also considerably less computational time, specially when $N$ is large as in the thermodynamical limit.

\par For our case of interest, individual dephasing, this formalism works even better as the entanglement contained in an arbritrary multipartition of a locally-dephased graph state is equivalent not to the average entanglement of an ensemble of smaller effective boundary systems but just one. Furthermore, such effective system is simply composed by the locally-dephased original system itself but without all non-boundary qubits. The key point behind this idea is that all the control-$Z$ gates that define $\ket{G_{0\ ...\ 0}}$ in Eq. \eqref{graph state} commute with the dephasing map \eqref{decoherence} \cite{Comment}. Hence, the order in which channels $\mathcal{E}_i$ and the control-Z gates are applied on the product state $\bigotimes_{k=1}^N \ket{+}_k$ to obtain $\Lambda\big(\ket{G_{0\ ...\ 0}}\bra{G_{0\ ...\ 0}}\big)$ is irrelevant. In particular,  state  \eqref{decoherence1} is also obtained if the control-$Z$ gates act after the dephasing channels. Now, all $CZ$'s not crossing any boundary are local unitary operations with respect to the multipartition (see Figs. \ref{linear cluster} and \ref{2Dcluster} for simple examples). Thus, because every entanglement quantifier is invariant under local unitary operations, as far as what concerns the amount of entanglement in the multipartition one can simply forget about these non boundary-crossing $CZ$'s.

\par In what follows we apply this idea to some well-known, paradigmatic examples of graphs to show that there exists a range of temperatures where the associated thermal graph state possesses multipartite bound entanglement. We do it first for the case of constant couplings $B_i\equiv B$ to transmit the essential idea clearly and then move to the arbitrary-coupling case.

\subsection{The linear cluster with equal couplings}
\label{OneD}

\begin{figure}
\begin{center}\includegraphics[width=1\linewidth]{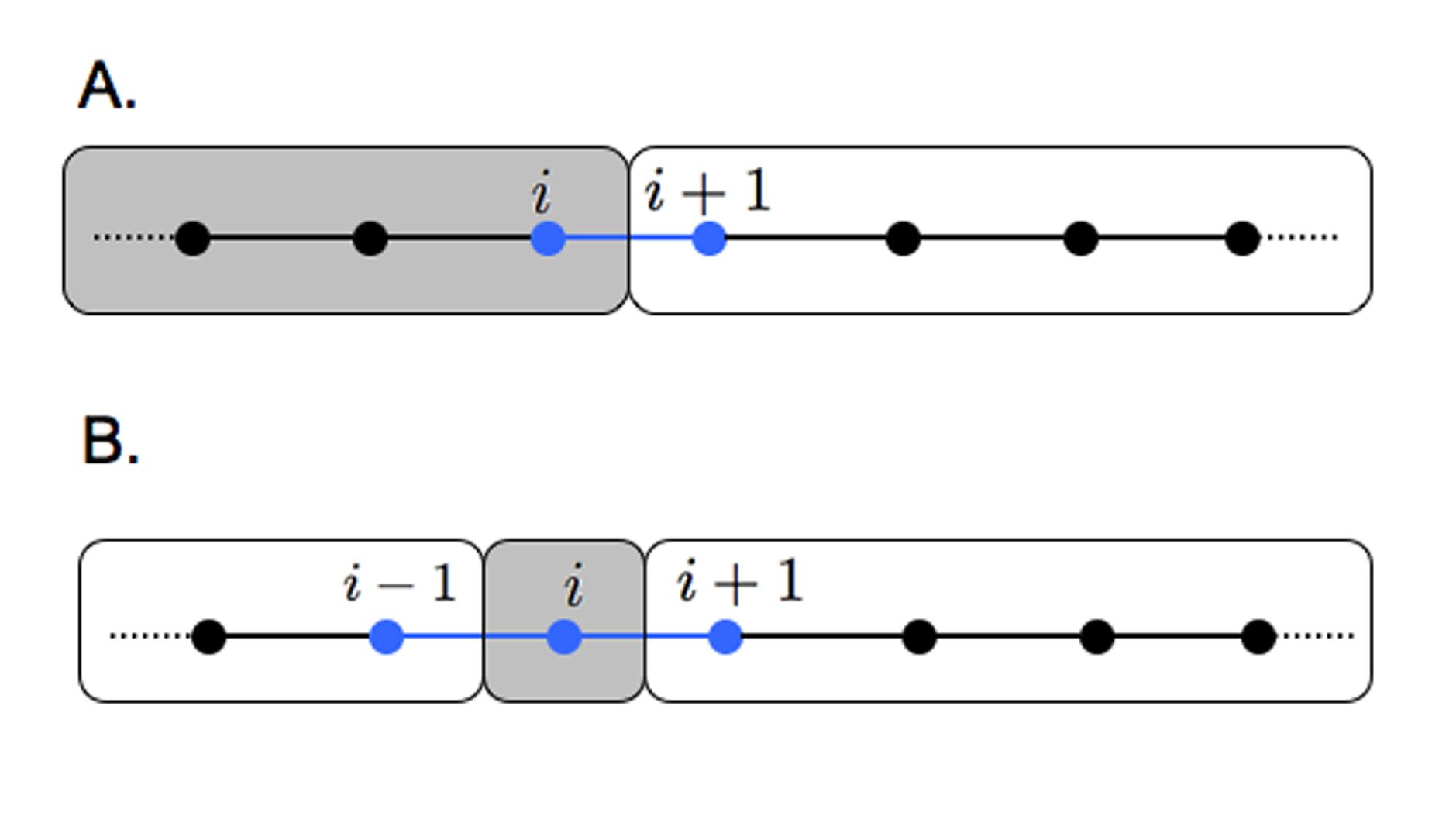}
\caption{Two possible bipartitions of the linear cluster: the system is split into two subpartitions, represented by the grey and white regions.  Note that all control-Z gates corresponding to black edges act locally with respect to the bipartition adopted, and thus do not change its entanglement.  {\bf A.} The systems is divided in two contiguous blocks of spins. The entanglement in this bipartition is equivalent to the entanglement between the two qubits in blue (see text also). {\bf B}. Another possible bipartition, this time non-contiguous blocks of spins are part of the same single subpartition (the white one). In this case all the entanglement is contained in the three-qubit boundary system shown in blue. }
\label{linear cluster}
\end{center}
\end{figure}

Let us start by the simplest example: the linear cluster state. Here the defyning graph $G$ is the linear graph sketched in Fig. \ref{linear cluster}. We denote its thermal state as ${\rho_{1D}}_T$. First we consider a bipartition of the system  into two contiguous blocks of spins (Fig. \ref{linear cluster}A), say from qubit $i$ to the left (grey) and from qubit $i+1$ to the right (white). We can easily see from the figure that all control-Z gates but one (in blue) act as local unitary operations, and thus have no effect on the entanglement in the bipartition considered. As a result, the entire entanglement between any two contiguous blocks of spins in ${\rho_{1D}}_T$ is equivalent to that of the simple two-qubit thermal graph state in boundary pair $i$-$i+1$. Then, by imposing the entanglement between $i$ and $i+1$ to vanish we can establish the critical temperature for which the entire thermal cluster is separable with respect to the bipartition under scrutiny. Any entanglement quantifier valid for two-qubit mixed states would do for this aim, so we choose the simplest one to calculate: the \emph{negativity} \cite{negat}. The negativity $Neg(\rho)$ of a state $\rho$ is the sum of the absolute values of the negative eigenvalues of $\rho^\Gamma$, where $\rho^\Gamma$ is the partial transposition of $\rho$ according to some bipartition. The negativity $Neg_{i|i+1}({\rho_{1D}}_T)$ of the thermal state of pair $i|i+1$ is readily calculated to be $Neg_{i|i+1}({\rho_{1D}}_T)=\frac{1}{4}(2-2p_i-2p_{i+1}+ p_ip_{i+1})$. For the case of constant couplings $p_i=p_{i+1}\equiv p=\frac{2}{1+e^{B/T}}$ condition $Neg_{i|i+1}({\rho_{1D}}_T)=0$ leads to the critical temperature  \cite{Kay}
\begin{equation}\label{Tc1}
{T^c}_{i|i+1}(B)=\frac{-B}{\ln(\sqrt{2}-1)}\approx 1.1 B.
\end{equation}

For $T\geq {T^c}_{i|i+1}(B)$, since the partition is separable, it is not possible to extract {\it any type of entanglement} between any two contiguous blocks joined at spins $i$ and $i+1$ from a thermal 1D graph state by applying contigous-block-local operations (arbitrary operations acting on spins 1 to $i$ and on spins $i+1$ to $N$). This imposes strong restrictions on the operations one needs to apply to ${\rho_{1D}}_T$ to distill some entanglement (if any) from some other of its multipartitions. For instance, if individual local operations on each spin are applied, no entanglement can be distilled from this bipartition of ${\rho_{1D}}_T$ if $T\geq{T^c}_{i|i+1}(B)$, for these are  a particular case of contigous-block-local operations. Now, since here we have taken all coupling strengths equal, the critical temperature for separability of contiguous blocks is the same for all $i$. This means that for $T\geq {T^c}_{i|i+1}(B)$ all contiguous blocks of ${\rho_{1D}}_T$ are in a separable state. So, no entanglement between any two particles can be extracted by LOCC, as for any two particles a contiguous-block bipartition can be found in which each particle lies on a different side of the partition and is therefore separable from the other. Ergo, {\it no entanglement at all} can be extracted  from ${\rho_{1D}}_T$ for  $T\geq {T^c}_{i|i+1}(B)$ by LOCC.

\begin{figure}
\begin{center}\includegraphics[width=1\linewidth]{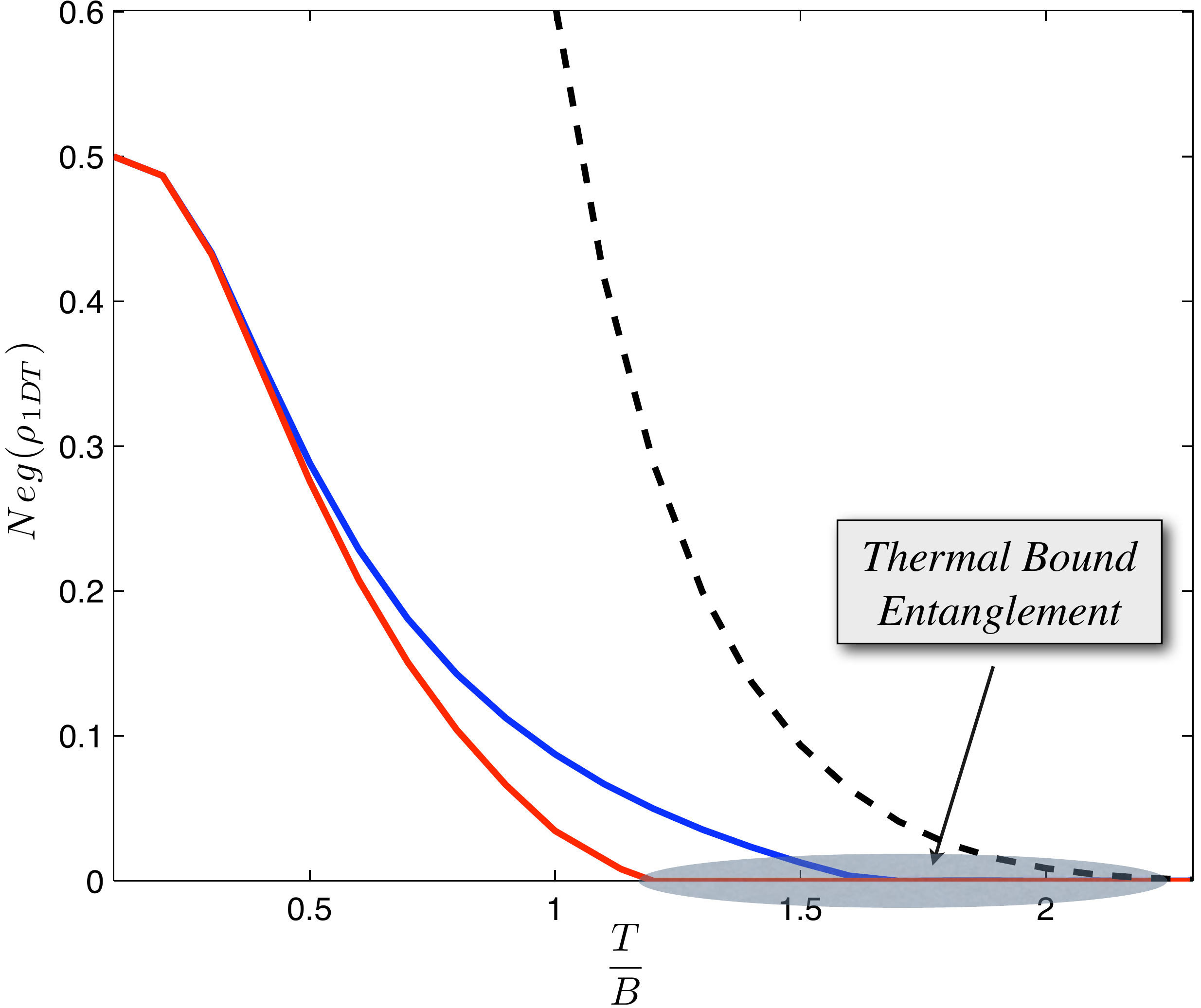}
\caption{Negativities of the thermal linear cluster state ${\rho_{1D}}_T$, as
a function of temperature $T/B$. The considered bipartitions are those of Fig. 1: Òany
two contiguous blocksÓ, in red and Òany qubit versus the
restÓ, in blue.   None of these curves depends on the size $N$ of the graph. Therefore such thermal bound entanglement is also present in the macroscopic thermodynamical limit. Besides, we also display in dashed line the negativity of the even-odd partition, where particles with even label belong to subsystem $A$ and particles with odd label to subsystem $B$.  In this case, the negativity does depend on the system's system and the plot is done for $N=12$. Our numerical investigations suggest that the even-odd partition is the most robust bipartition, \ie the one with the highest critical temperature of vanishing negativity.  The shadded region shows the range of temperatures where the thermal system possesses bound entanglement (see text).}
\label{negativity}
\end{center}
\end{figure}

We now find another (non-contiguous) family of bipartitions for which the separability temperature, ${T^c}_{i|i-1,i+1}(B)$, is strictly larger than ${T^c}_{i|i+1}(B)$. This suffices to prove the non-separability of ${\rho_{1D}}_T$ -- and therefore the presence of multipartite bound entanglement in --  for a range of temperatures ${T^c}_{i|i+1}(B)\leq T \leq{T^c}_{i|i-1,i+1}(B)$ \footnote{Note that in the first version of    arXiv:0902.4343 it was erroneously claimed that ${T^c}_{i|i+1}(B)$ is the separability temperature for $\rho_{{1D}_T}$ \cite{private}.}.  Consider for instance the entanglement between the $i$-th qubit in the chain and all the rest (see Fig. \ref{linear cluster} B). In this case, one can ignore all but two control-Z gates in the calculation of the entanglement. So, for these partitions, the entanglement (again quantified here by the negativity) of ${\rho_{1D}}_T$ is equivalent to that of the central particle vs. its two neighbours in a linear thermal cluster state of only three qubits. This negativity vanishes at a critical temperature ${T^c}_{i|i-1,i+1}(B)$ that turns out to be strictly larger than  ${T^c}_{i|i+1}(B)$ (${T^c}_{i|i-1,i+1}(B)\gtrsim1.6 B$, see Fig. \ref{negativity}).  In the range ${T^c}_{i|i+1}(B)\leq T \leq{T^c}_{i|i-1,i+1}(B)$, even though ${\rho_{1D}}_T$ possesses entanglement, we already know that none of it can be distilled through local operations assisted by classical communication. Such entanglement can only be extracted if particles $i-1$ and $i+1$ interact, which is of course not a local operation. Therefore, in this range of temperatures thermal cluster state ${\rho_{1D}}_T$ possesses multipartite bound entanglement.

\par It is important to stress that none of the latter results or conclusions depends at all on the size $N$ of the graph This allows us to guarantee that thermal bound entanglement is also present in macroscopic specimens of these graphs. This independence of the entanglement on the graph's size is precisely the key point behind the method we used to simplify the calculation of the negativity of ${\rho_{1D}}_T$. In general, in order to calculate the negativity of an $N$-qubit mixed state,  one would need to diagonalize a $2^N\times 2^N$ matrix, which  already for a few tens of qubits cannot even be written down by a current classical computer. In the example studied in this subsection, this method has enabled us to obtain results for arbitrarily large systems calculating only negativities of two-qubit and three-qubit systems.

\subsection{The 2D square cluster with equal couplings}
\label{TwoD}

\begin{figure*}
\begin{center}\includegraphics[width=0.7\linewidth]{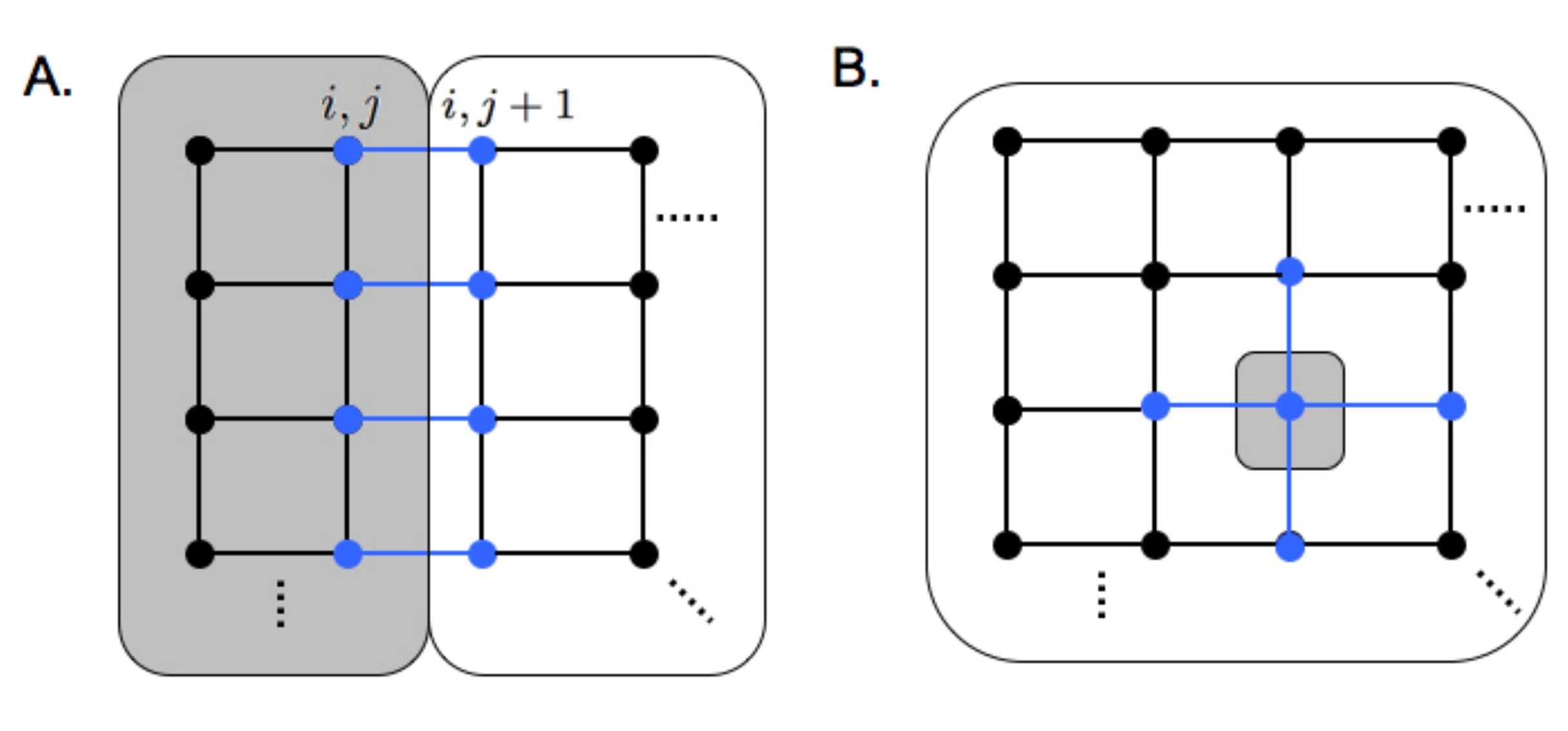}
\caption{Two different bipartitions of a 2D cluster. {\bf A.} The system is divided vertically into two parts. The only control-Z gates not acting as local unitary operations are the boundary-crossing ones, shown in blue. Thus the entanglement in any such bipartition of the  2D cluster in a thermal state is equivalent to that  of several copies of a two-qubit thermal cluster state (represented in blue). {\bf B.} The cluster is partitioned into a central spin and all other ones. The entanglement is this time equivalent  to that  between the central spin and only its four neighbours in a five-qubit thermal graph state in a star configuration (also in blue). }\label{2Dcluster}
\end{center}
\end{figure*}

Our next example is the 2D thermal cluster state ${\rho_{2D}}_T$, for which the associated graph $G$ is a $\sqrt{N}\times\sqrt{N}$ square lattice (see Fig. \ref{2Dcluster}). The columns of $G$ are labeled by index $1\leq i\leq\sqrt{N}$ and the rows by index $1\leq j\leq\sqrt{N}$. At $T=0$ this state is known to be a universal resource for one-way quantum computation. As a consequence, understanding the entanglement properties of this model under realistic noisy conditions is of course very important from a practical point of view.

To prove the existence of thermal multipartite bound entanglement in this example we follow exactly the same reasoning as in the previous subsection: establish a region of temperatures  where the system is separable with respect to any two contiguous blocks -- and therefore nondistillable with respect to (contiguous-block) local operations while still being entangled in other partitions.  In Fig. \ref{2Dcluster} A we consider a vertical partition of the system from column $i$ to the left (grey) and from column $i+1$ to the right (white). Following the same steps as before, the calculation of entanglement in this bipartition can be reduced to that of a product of the $\sqrt{N}$ two-qubit thermal cluster states lying on the boundary. In this way we see that the critical temperature for separability (of all contiguous blocks) is again given by \eqref{Tc1}. Once again we consider  another family of bipartitions, say, any qubit $ij$ inside the lattice ($1< i<\sqrt{N}$ and $1< j<\sqrt{N}$) vs. all the rest of the qubits (see Fig. \ref{2Dcluster} B). The entanglement in this bipartition is equivalent to that between the qubit $ij$ and its four neighbours,  in a thermal graph state in a star configuration. The temperature for which the negativity vanishes in this case happens to be ${T^c}_{ij|i-1,i+1,j+1,j-1}(B)\approx 2.5B$, which is again strictly larger than ${T^c}_{i|i+1}(B)$.

\subsection{Unequal Hamiltonian couplings}

Consider now arbitrary couplings $B_i$ in Hamiltonian \eqref{H} with average value $B$, that is: $\frac{1}{N}\sum_{i=1}^{N}B_i=B$. We first restrict ourselves to the familiar one-dimensional graph studied in subsection \ref{OneD}, as this already qualitatively  captures all the essential  changes that can appear when different couplings  are present. The contigous-block separability condition $Neg_{i|i+1}({\rho_{1D}}_T)=\frac{1}{4}(2-2p_i-2p_{i+1}+ p_ip_{i+1})=0$ now involves $p_i\equiv\frac{2}{1+e^{B_i/T}}\neq p_{i+1}\equiv\frac{2}{1+e^{B_{i+1}/T}}$. Thus the critical temperature is implicitly expressed by the equation
\begin{equation}
\label{Tcunequal}
1=e^{-\frac{B_i}{T}}+e^{-\frac{B_{i+1}}{T}}+e^{-\frac{B_i+B_{i+1}}{T}}.
\end{equation}
Equation \eqref{Tcunequal} above has a unique real solution $T={T^c}_{i|i+1}(B_i,B_{i+1})$, but is however non-invertible in general. Clearly, critical temperature ${T^c}_{i|i+1}(B_i,B_{i+1})$ is symmetric under the exchange of $B_i$ and $B_{i+1}$. Also, numerical inspections immediately show that it is a monotonously growing function of $B_i$ ($B_{i+1}$) for fixed $B_{i+1}$ ($B_i$). Furthermore, under some contraints, as for example $B_i+B_{i+1}$ constant, the critical temperature is maximal when the couplings are equal.

As before, we compare ${T^c}_{i|i+1}(B_i,B_{i+1})$ with the critical temperature ${T^c}_{i|i-1,i+1}(B_{i-1},B_{i},B_{i+1})$ corresponding to the bipartition which distinguishes qubit $i$ versus the rest. For all cases we have studied, ${T^c}_{i|i-1,i+1}(B_{i-1},B_{i},B_{i+1})$  turns out to be strictly larger than  ${T^c}_{i|i+1}(B_{i},B_{i+1})$ $\forall\ 1\leq i\leq N$. If this is true in general, it implies that neither the appearance, nor the range of temperatures, of bound entanglement will be considerably affected by small deviations of $B_{i}$ from the mean value $B$.

\par However, the critical temperatures can be considerably sensitive to the values of the coupling constants at each site $i$ when the deviations are large (say, of the order of $B$ itself). In such case, it might as well happen  that ${T^c}_{i|i+1}(B_{i},B_{i+1})<{T^c}_{i|i-1,i+1}(B_{i-1},B_{i},B_{i+1})<{T^c}_{k|k+1}(B_{k},B_{k+1})<{T^c}_{k|k-1,k+1}(B_{k-1},B_{k},B_{k+1})$, for some $i\neq k$. Remarkably, also in these cases the system displays a finite range of temperatures for which it is bound-entangled. To see this, it suffices to consider the site $i_{max}$ for which the critical temperature of contigous-block separability is the largest of all, {\it i e}. ${T^c}_{i_{max}|i_{max}+1}(B_{i_{max}},B_{i_{max}+1})\geq{T^c}_{i|i+1}(B_{i},B_{i+1})$, $\forall\ 1\leq i\leq N$. By the same reasonings as before, from $T={T^c}_{i_{max}|i_{max}+1}(B_{i_{max}},B_{i_{max}+1})\equiv{T^c}_{i_{max}|i_{max}+1}$ on ${\rho_{1D}}_T$ is non-distillable, but it is entangled up to  $ T={T^c}_{i_{max}|i_{max}-1,i_{max}+1}(B_{i_{max}-1},B_{i_{max}},B_{i_{max}+1})\equiv{T^c}_{i_{max}|i_{max}-1,i_{max}+1}$. Therefore (assuming again that ${T^c}_{i_{max}|i_{max}-1,i_{max}+1}>{T^c}_{i_{max}|i_{max}+1}$  is always true), for $ T\in[{T^c}_{i_{max}|i_{max}+1},{T^c}_{i_{max}|i_{max}-1,i_{max}+1})$ the state is bound-entangled. The main conclusion to draw from the considerations in this paragraph is that rather than a peculiarity of the  (ideal) case of equal coupling strengths, the presence of bound entanglement in a finite range of temperature appears to be a {\it general phenomenon}  for (arbitrarily large) thermal graph states

Finally, let us stress that similar results hold also for other graphs with different couplings. There again different partitions will give rise to different critical temperatures implying again the presence of bound entanglement along the lines of Sec. \ref{TwoD}.


\section{Conclusion}

Considering thermal graph states we have shown the presence of bound entanglement in systems containing a macroscopic number of spins with finite range interactions. This result extends previous results \cite{Oscillators} for bosonic chains to fermionic systems. Our findings suggest that thermal bound entanglement could manifest also in more general systems, since this seems to be a robust feature against, in particular, to modifications in the symmetries of the Hamiltonian.

In this paper we considered systems with three (or more) body
interactions. Thus a natural question arises regarding the
presence of macroscopic thermal bound entanglement also in spin
models containing only $2$-body interactions. This was indeed
found in the case of harmonic oscillators systems
\cite{Oscillators}. Actually, we expect thermal bound entanglement
to be a common phenomenon of general many-body systems since it is
very unlikely that the negativities of all possible bipartitions
of a system vanish at the same temperature. However, proving that
for systems in the thermodynamical limit turns to be nontrivial.

It is also worth mentioning that we used the negativity of some
bipartitions of the system to detect inseparability of the thermal
state. So our method can not detect bound entanglement in the case
that all bipartitions of a system have positive partial
transpositions. This was in turn found in the case of small spin
systems \cite{Toth}. We leave the existence of such kind of bound
entanglement in macroscopic systems as an open problem.

An interesting question concerns the utility of thermal graph states for information processing. For the region of temperatures such that these states are distillable, one could in principle apply first a distillation protocol (\eg see \cite{Kay}) before using the state as a resource for quantum information. However, for the regions of bound entanglement, not even this experimentally-demanding strategy would work.

Since in any practical implementation the temperature is  always non-null, thermal bound entangled  graph states are ideal probes to explore the limitations of realistic (experimentally feasible) measurement-based quantum computation. This question is certainly of great interest and can be the subject of further analysis.

\section{Acknowledgements}

We would like to thank M. Hajdusek and V. Vedral for useful discussions. We acknowledge the European QAP, COMPAS and PERCENT projects, the Spanish MEC FIS2007-60182 and Consolider-Ingenio QOIT projects, the ``Juan de la Cierva'' grant, and the Generalitat de Catalunya, for financial support.

\appendix
\section{Thermal graph state  as a dephased graph state for arbitrary couplings}
\label{App}
Thermal state \eqref{rho thermal} expressed in the eigenbasis of Hamiltonian \eqref{H}, $\{\ket{G_{\mu_1\ ...\ \mu_N}}\}$, reads  (disregarding its normalization) $\rho_{T}\equiv e^{-H/ T}=\sum_{\mu_1\  ...\ \mu_N=0}^1e^{\frac{1}{2T}\sum_{i=1}^N B_i (-1)^{\mu_i}}\ket{G_{\mu_1\ ...\ \mu_N}}\bra{G_{\mu_1\ ...\ \mu_N}}$. Next we explicitly evaluate dephased state $\Lambda\big(\ket{G_{0\ ...\ 0}}\bra{G_{0\ ...\ 0}}\big)$ using Eqs. \eqref{decoherence} and \eqref{dephasing} :
\begin{widetext}
\begin{eqnarray}
\label{decoherenceApp}
\nonumber
\Lambda\big(\ket{G_{0\ ...\ 0}}\bra{G_{0\ ...\ 0}}\big)\equiv\mathcal{D}_1\otimes ...\ \mathcal{D}_N\big(\ket{G_{0\ ...\ 0}}\bra{G_{0\ ...\ 0}}\big)
\equiv\sum_{\mu_1\  ...\ \mu_N=0}^1\prod_{i=1}^N(1-p_i/2)^{|\mu_i-1|_2}(p_i/2)^{|\mu_i|_2}\ket{G_{\mu_1\ ...\ \mu_N}}\bra{G_{\mu_1\ ...\ \mu_N}},
\end{eqnarray}
\end{widetext}
where ``$|\ |_2$" stands for ``modulo 2".  The latter is equal to the former expression for $\rho_{T}$  if and only if each and all of the terms in the summation are equal. That is, if and only if $\prod_{i=1}^N(1-p_i/2)^{|\mu_i-1|_2}(p_i/2)^{|\mu_i|_2}\equiv e^{\frac{1}{2T}\sum_{i=1}^N B_i (-1)^{\mu_i}}$, which  when Eq. \eqref{probability} holds --and using the fact that $\frac{1}{1+e^{B_i/T}}\equiv\frac{e^{-B_i/2T}}{e^{-B_i/2T}+e^{B_i/2T}}$ and  $1-\frac{1}{1+e^{B_i/T}}\equiv\frac{e^{B_i/2T}}{e^{-B_i/2T}+e^{B_i/2T}}$ -- in turn reads $\prod_{i=1}^N\frac{e^{(-1)^{\mu_i}B_i/2T}}{e^{-B_i/2T}+e^{B_i/2T}}\equiv e^{\frac{1}{2T}\sum_{i=1}^N B_i (-1)^{\mu_i}}$ which can in turn be immediately checked to be true up to a constant normalization factor.



\begin{thebibliography}{20}
\bibitem{Exception}
Exceptions are of course the Hamiltonians associated to the maximally connected graphs (that give rise to GHZ states \cite{GHZ}), which are never local; and the ones associated to  expander graphs, which can be local but not necessarily of finite range.
\bibitem{Comment} This was indeed the key point behind the results of \cite{Kay,Cavalcanti,HeinDurBrie}.
\end{thebibliography}
\end{document}